\documentclass[prd,superscriptaddress,twocolumn,showpacs,amsmath,
preprintnumbers,showkeys]{revtex4}

\usepackage{graphicx}
\usepackage{dcolumn}
\usepackage{bm}

\usepackage{axodraw4j}

\def\beq{\begin{equation}}
\def\eeq{\end{equation}}
\def\beeq{\begin{eqnarray}}
\def\eeeq{\end{eqnarray}}

\def\LQCD{\Lambda_{\mbox{\rm\scriptsize QCD}}}
\def\as{\alpha_{\mbox{\rm\scriptsize s}}}
\def\lrang#1{\left\langle#1\right\rangle}
\def\cO#1{{\cal{O}}\left(#1\right)}

\begin{document}
\title{pQCD physics of multiparton interactions }
\author{ B.\ Blok}
\affiliation{Department of Physics, Technion---Israel Institute of
Technology, 32000 Haifa, Israel} \email{blok@physics.technion.ac.il}
\author{Yu.\ Dokshitzer }
\affiliation{Laboratory of High Energy Theoretical Physics
(LPTHE), University Paris 6, Paris, France \footnote{On leave of
absence: St.\ Petersburg Nuclear Physics Institute, 
Gatchina, Russia}} \email{yuri@lpthe.jussieu.fr}
\author{L.\ Frankfurt} \affiliation{School of Physics and
Astronomy, Raymond and Beverly Sackler Faculty of Exact Sciences,
Tel Aviv University, 69978 Tel Aviv, Israel} \email{frankfur@tauphy.tau.ac.il}
\author{ M.\ Strikman}
\affiliation{Physics Department, Penn State University, University
Park, PA, USA} \email{strikman@phys.psu.edu}

\begin{abstract}

We study production of two pairs of jets in 
hadron--hadron collisions 
in view of extracting contribution of {\em double hard interactions}\/ of three and four partons ($3\to4$, $4\to4$).
Such interactions, in spite of being power suppressed at the level of the total cross section,  
become comparable with the standard hard collisions of two partons, $2\to4$, 
in the {\em back-to-back kinematics}\/ when the transverse momentum imbalances of two pairing jets are relatively small.

We express differential and total cross sections for two-dijet production in double parton collisions
through the generalized two-parton distributions, $_2$GPDs \cite{BDFS1}, 
that contain large-distance two-parton correlations of non-perturbative origin
as well as small-distance correlations due to parton evolution.
We find that these large- and small-distance correlations participate in different
manner in 4-jet production, and treat them in the leading logarithmic approximation 
of pQCD that resums collinear logarithms in all orders.

A special emphasis is given to $3\to4$ double hard interaction processes that occur
as an interplay between large- and short-distance parton correlations and were 
not taken into consideration by approaches inspired by the parton model picture.
We demonstrate that the $3\to4$ mechanism, being of the same order in $\as$ as the $4\to4$ process,
turns out to be {\em geometrically enhanced}\/ compared to the latter and 
should contribute significantly to 4-jet production.

The framework developed here takes into systematic consideration perturbative $Q^2$ evolution of $_2$GPDs.  
It can be used as a basis for future analysis of NLO corrections to multi-parton interactions (MPI) 
at LHC and Tevatron colliders, in particular for improving evaluation of QCD backgrounds to new physics searches.
\end{abstract}

\maketitle
\setcounter{page}{1}
\section{Introduction}

Understanding the rates and the structure of multi-jet production in hadron--hadron collisions
is of primary importance for new physics searches.

Production of high transverse momentum jets is a hard process which implies a head-on collision of QCD partons
--- quarks and/or gluons --- from the small-distance wave functions of initial hadrons.
Cross section of a hard collision is small compared with the size of hadron, $\sigma\propto Q^{-2}\ll R^2_{\mbox{\scriptsize hadr}}$,
with $Q^2$ the scale related to transverse momenta of the produced jets, $Q^2\sim j_\perp^2$.
Therefore, typically it is two partons that experience a hard collision in a given event.
A large angle scattering of these two partons produces two (or more) final state partons
 that manifest themselves as hadron large transverse momentum jets.
At the same time, one cannot exclude a possibility that more than one pair of partons
 happen to collide in a given event, giving rise to a multi-jet event.
A possibility of a double hard collision becomes more important with increase of
 the energy of the collision where scattering off small $x$ partons which have much higher densities becomes possible.

In recent years multiparton collisions have attracted close attention.
Following the pioneering work of Refs.~\cite{TreleaniPaver,mufti},
a large number of related theoretical papers appeared
\cite{Treleani,Wiedemann,Frankfurt,Frankfurt1,SST}, based on the parton model and geometrical picture in the impact parameter space.
More recently, this topic has been intensively discussed in view of the LHC program \cite{Perugia,Fano}.
Monte Carlo  event generators that produce multiple parton collisions are being developed \cite{Pythia,Herwig,Lund}; theoretical
papers exploring properties of  double parton distributions and discussing their QCD evolution have appeared
\cite{BDFS1,Diehl,DiehlSchafer}.

In our view, however, important elements of QCD that are necessary for theoretical understanding
of the multiple hard interactions issue have not yet been properly taken into account by above-mentioned intuitive approaches.

The problem is, sort of, educational: both the probabilistic picture, the MC generator technology is based upon,
and the familiar Feynman diagram technique, when used in the momentum space, prove to be inadequate
for careful analysis and understanding of the physics of multiple collisions.

From experience gained by treating standard (single) hard processes, one became used to a motto that 
a large momentum transfer scale $Q^2$ ensures the dominance of small distances, $r^2\sim Q^{-2}$, in a process under consideration.
With the multiple collisions under focus, however, one has to distinguish two space-time scales:
that of {\em localization}\/ of the parton participating in a hard interaction, $\Delta r^2 \sim Q^{-2}$,
and that of {\em transverse separation}, $\Delta\rho$, between the two hard collision vertices.
The latter can be large, of the order of the hadron size, even for large $Q^2$.

In order to be able to trace the relative distance between the partons, one has to use the mixed longitudinal
momentum--impact parameter representation which, in the momentum language, reduces to introduction
of a mismatch between the transverse momentum of the parton in the amplitude and that of the same parton
in the amplitude conjugated.

\medskip
Another unusual feature of the multiple collision analysis that may look confusing at the first sight is the fact that
--- even at the tree level --- {\em the amplitude}\/ describing the double hard interaction process contains additional integrations
over longitudinal momentum components; more precisely --- over the difference of the (large) light-cone momentum components
of the two partons originating from the same incident hadron (see Section~\ref{SubSec:PT34}).

\bigskip

In the previous short publication \cite{BDFS1} we have considered production of two pairs of {\em nearly back-to-back}\/
jets resulting from simultaneous hard collisions of two partons from the wave function of one incident hadron
with two partons from the other hadron (``four-to-four'' processes).
As we have shown, this necessitates introduction of a new object --- a generalized double parton distribution, $_2$GPD,
that depends on a new transverse momentum parameter $\vec\Delta$ conjugate to the relative distance between
the two partons in the hadron wave function.
Generalized double parton distributions provide a natural framework for incorporating longitudinal and transverse correlations
between partons in the hadron wave function at the $Q_0^2$ scale, and for tracing the perturbative $Q^2$-evolution of the correlations.

The corresponding 4-jet cross section can be expressed in terms of $_2$GPD's as follows
\beq\label{eq:1}
\begin{split}
& \frac{d\sigma(x_1,x_2,  x_3,x_4)}{d\hat{t}_1\,d\hat{t}_2}
 = \>\frac{d\sigma^{13}}{d\hat{t}_1}\, \frac{d\sigma^{24}}{d\hat{t}_2}\cr
& \quad \times  \int\frac{d^2\vec{\Delta}}{(2\pi)^2}\, D_a(x_1,x_2;\vec{\Delta})\, D_b(x_3,x_4;-\vec{\Delta}).
\end{split}
\eeq
The factor on the second line has dimension of inverse area:
 \beq
\frac{1}{\pi R^2_{\rm int}}=\int
 \frac{d^2\overrightarrow{\Delta}}{(2\pi)^2}\frac{D(x_1,x_2,\overrightarrow{\Delta})
 D(x_3,x_4,-\overrightarrow{\Delta})}{D(x_1)D(x_2)D(x_3)D(x_4)},
 \label{b3}
 \eeq
 where $D(x_i)$ are the corresponding one-parton distributions. 
 The ratio of the product of two single-inclusive cross sections and the double-inclusive cross section ($\pi R_{int}^2$) 
 is often referred to in the literature as ``an effective cross section'' $\sigma_{\mbox{\scriptsize eff}}$.  
 We prefer, however, not to look at this quantity as a cross section, since it reflects transversal area of parton overlap 
 as well as longitudinal correlations of the partons. 
 At the same time, it has little to do with the measure of the strength of the interaction, 
 which is what ``cross section'' represents.

In a two-parton collision, scattered partons form two nearly back-to-back jets, while additional jets (should there be any)
tend to be softer and to align with the directions of initial and final partons, because of collinear enhancements
due to radiative nature of secondary partons.
Such will be typical characteristics of a {\em 4-jet event}, in particular.
On the other hand, four jets produced as a result of a {\em double hard collision}\/ of two parton pairs would, on the contrary,
form {\em two pairs}\/ of {\em nearly back-to-back}\/ jets.
This kinematical preference is in stark contrast with  ``hedgehog-like'' configurations
of four jets stemming from a single collision and can be used in order to single out double hard collisions experimentally.

Such experimental studies were recently carried out by the CDF and D0 collaborations
who have studied production of three jets + photon \cite{Tevatron1,Tevatron2,Tevatron3,Perugia}.
The analysis of the data performed in \cite{Frankfurt,Frankfurt1} using information about generalized parton distribution (GPDs)
obtained from the study of hard exclusive processes at HERA has found that  observed $3~jet +\gamma$ rates
were a factor $\ge 2$ higher than the expected rates based on a naive model that neglected correlations
between partons in the transverse plane.

The use of $_2$GPD allows one to incorporate such correlations and predict their $Q^2$ evolution.

\bigskip
On the theory side, the back-to-back enhancement has been discussed, at tree level, in a number of  studies of various channels
(see, for example, discussion of the 2~jets+$b\bar{b}$ in \cite{Berger} and references therein).

In the present paper we study perturbative radiative effects in the {\em differential}\/ 4-jet distribution in the back-to-back
kinematics and derive the expression for the corresponding cross section in the leading logarithmic collinear approximation.
It takes into account QCD evolution of the generalized double parton distributions as well as effects due to multiple
soft gluon radiation, and turns out to be a direct generalization of the known ``DDT formula''
for back-to-back production of two large transverse momentum particles in hadron collisions \cite{DDT}.

We also discuss and treat new specific correlations between transverse momenta of jets
due to 3-parton interactions producing 4 jets,``three-to-four''.
Such processes are induced by perturbative splitting of a parton from one of the hadrons, the offspring of which enter 
double hard collision with two partons from the wave function of the second hadron. The hard scale of this parton splitting 
is determined by transverse momentum imbalances of pairs of jets, $\delta_{13}$, $\delta_{24}$, and exhibits specific collinear 
enhancement in the kinematical region where two jet imbalances practically compensate one another, 
$\delta'^2=(\vec\delta_{13}+\vec\delta_{24})^2\ll \delta_{13}^2\simeq\delta_{24}^2$.

Consistently taking into account three-to-four parton process solves a longstanding problem 
of double counting in treating multi-parton interactions. 

\medskip
Discussion of the 2-parton distribution has a long history.
It is commonly defined in the momentum space as a 2-particle inclusive quantity depending on two parton momenta, 
see \cite{Kirschner,Snigirev}. Being related to (the imaginary part of) a certain {\em forward}\/ scattering amplitude, 
it therefore disregards impact parameter space geometry of the interaction.
Exploring properties of 2-parton distributions so defined, an approach to the study of the multiple jet production 
has been recently suggested in Ref.~\cite{stirling}.
The reason why this approach has faced difficulties, \cite{stirling1}, and did not solve, in our view, 
the problem of systematic pQCD analysis of 4-jet production is clear:
it did not incorporate effects due to variations of the transverse separation between the partons ---
information encoded by $_2$GPD's but not by the 2-parton momentum distributions.

The $_2$GPD's were recently used in Ref.~\cite{Ryskin} for intuitive description of the total 4-jet production cross section.
However, the differential distributions were not discussed in that paper, and not all relevant pQCD contributions were included,
so that our results are different from the ones obtained in \cite{Ryskin}.

\bigskip
The paper is organized as follows.

In Section \ref{PTAnalysis} we recall the main ingredients of the perturbative analysis  based on selection of maximally 
collinear enhanced contributions in all orders.
In Section \ref{Sec:EvEq} we present the evolution equation for generalized two-parton distributions.
Section \ref{Sec:PT} is devoted to the perturbative analysis of small-distance correlations between partons.
The main result of the paper --- the differential distribution of 4-jet production in the back-to-back kinematics --- is
 formulated in Section~\ref{Sec:4324}, and the total cross section of two-dijet production is described in Section~\ref{total}.
Conclusions and outlook are presented in Section~\ref{Conclude}.

\section{Perturbative Analysis\label{PTAnalysis}}
\subsection{Hard scales}
The perturbative approach implies that all hardness (transverse momentum) scales that characterize
 the problem are comfortably larger than the intrinsic QCD scale $\LQCD$: $Q_i^2\gg\LQCD^2$.
The process under consideration may have up to five hard scales involved.

Indeed, in the leading order in $\as$, large transverse momentum partons are produced in pairs and have nearly
 opposite transverse momenta, setting the hard scale $Q_1^2=j_{1\perp}^2\simeq j_{3\perp}^2$.
Within the parton model framework (neglecting finite smearing due to intrinsic transverse momenta of incident partons),
 one has $d\sigma\propto \delta(\vec{j}_{1\perp}+\vec{j}_{3\perp})$.
Secondary QCD processes ---  evolution of initial parton distributions and accompanying soft
 gluon radiation --- introduce transverse momentum imbalance: $\vec{j}_{1\perp}+\vec{j}_{3\perp}=\vec{\delta}_{13}$ which
 constitutes another hard scale: $Q_1^2\gg \delta_{13}^2\gg \LQCD^2$. For production of four jets
 in the back-to-back kinematics, this gives four different hard scales. As we shall see below, in the 3-partons
 collisions producing four jets yet another scale enters the game:
$\delta'^2\gg\LQCD^2$ with $\vec{\delta'}= \vec{\delta}_{13}+  \vec{\delta}_{24}$ --- the total
 transverse momentum of the 4-jet ensemble.

In what follows we consider transverse momenta of all four jets to be of the same
 order, $Q^2=j_{1\perp}^2\simeq j_{3\perp}^2 \>\sim\> j_{2\perp}^2\simeq j_{4\perp}^2$. This
 is not necessary but helps to avoid complications in the hierarchy of relevant scales.

Finally, let us mention that  in what follows it will be tacitly implied that fixing
 these scales --- from the largest one, $Q^2$, down to smaller ones, $\delta$ and $\delta'$, --- is not
 compromised by uncertainties in determination of the transverse momenta of the jets.

\subsection{Back-to-back kinematics}

The basic 2-jet production cross section scales, asymptotically, as
\beq\label{eq:22}
   \frac{d\sigma^{(2\to2)}}{d\hat{t}} \propto \frac{\alpha_s^2}{Q^4}   .
\eeq
According to \eqref{eq:1}, production of four jets in simultaneous hard collisions of four partons yields
\begin{subequations}
\beq\label{eq:44}
 \frac{d\sigma^{(4\to4)}}{d\hat{t}_1d\hat{t}_2}  \>\propto \> R^{-2}\cdot \left(\frac{\alpha_s^2}{Q^4}\right)^2 \propto
 \frac{\alpha_s^4}{R^2\, Q^8},
\eeq
with $R^2=1/\lrang{\Delta^2}$ the characteristic distance between the two partons in the hadron wave function.
At large $Q^2$ this cross section is parametrically smaller than that for production of four well
 separated jets with transverse momenta $j_{i\perp}^2\sim Q^2$ in a 2-parton collision:
\beq\label{eq:24}
   \frac{d\sigma^{(2\to4)}}{d\hat{t}_1d\hat{t}_2} \>\propto  \> \frac{\alpha_s^4}{Q^6}
\eeq
\end{subequations}
(with transverse momenta of two out of four jets being integrated over).

Qualitatively, the production mechanism \eqref{eq:44} can be labelled a ``higher twist effect''.
Nevertheless, it may turn out to be essential --- comparable with the ``leading twist'' $2\to4$, Eq.~\eqref{eq:24} 
--- if one looks at specific kinematics of the 4-jet ensemble.

Let $z$ be the direction of colliding hadron momenta. Imagine that we are triggering on two  jets moving along the $x$ and $y$
 axes in the transverse plane, and look for two accompanying jets  inside some solid
 angles $\Delta\Omega\ll 4\pi$ around the $-x$ and $-y$ directions.
  The production mechanism \eqref{eq:24} does not populate this region: the higher order $2\to4$ QCD
 matrix element is enhanced when two final state partons become {\em quasi-collinear}\/ but is perfectly smooth
 in the back-to-back kinematics. Therefore, its contribution will be suppressed,
\[
 \left(\frac{\Delta\Omega}{4\pi}\right)^2 \quad\mbox{vs.}\quad \frac{1}{R^2Q^2} ,
\]
contrary to the $4\to4$ production mechanism \eqref{eq:44} which is concentrated in this very kinematical region.


\subsection{Collinear approximation}

The differential 4-jet production cross section possesses two collinear enhancements.
Depending on the kinematics of the jets, they are, symbolically,
\begin{subequations}\label{eqs:4434}
\beeq\label{eq:443444}
   d\sigma^{(4\to4)} &\propto& \frac{\as^2}{\delta_{13}^2\,\delta_{24}^2}\, d^2j_{3\perp}d^2j_{4\perp} \cdot d\Sigma, \nonumber\\
  && \quad \delta_{13}^2 \ll Q^2,  \quad \delta_{24}^2 \ll Q^2; \\
\label{eq:443434}
   d\sigma^{(3\to4)} &\propto& \frac{\as^2}{\delta'^2\,\delta^2} \> d^2j_{3\perp}d^2j_{4\perp} \cdot d\Sigma, \nonumber \\
   && \quad \delta'^2\ll \delta^2 \ll Q^2, \>  \delta^2= \delta_{13}^2\simeq \delta_{24}^2 .
\eeeq
\end{subequations}
Here $d\Sigma=d\Sigma(\hat{t}_1,\hat{t}_2)$ is the cross section integrated over the transverse momenta of the ``backward''  jets $3$ and $4$.
The integrated cross section $d\Sigma$ contains the squared matrix element of the four-parton production and is of the order
of ${\as^4}$, cf.\ Eq.~\eqref{eq:44}.
At the Born level,  the jets in pairs are exactly back-to-back, so that
 $d\sigma^{(4\to4)} \propto \delta^2(\vec{\delta}_{13})\delta^2(\vec{\delta}_{24})$ in \eqref{eq:443444}. 
 To have a  non-zero value of the transverse momentum imbalance,  one has to have additional large transverse momentum parton(s) produced.

In the second important contribution to the cross section, Eq.~\eqref{eq:443434}, one power of the coupling emerges from
 the splitting of a parton from one of the incident hadrons into two, and the second power
 is due to production of an additional final state parton with $\vec{k}_\perp = -\vec{\delta}'$.

In both cases the smallness due to additional powers of the coupling is compensated by two broad
(logarithmic) integrations over transverse momentum imbalances as indicated in \eqref{eqs:4434}.

\subsection{Double Logarithmic parton form factors}

In the leading order in $\as$, it suffices to have just one parton present with $\vec{k}_\perp=-\vec{\delta}_{13}$ in order to assure $\delta_{13}\neq0$.
At the same time, inclusive production of accompanying partons with transverse momenta $k_\perp$ turns out
 to be {\em suppressed}\/ in a broad interval $\delta_{13}^2\ll k^2_\perp\ll Q^2$, as long as one wants
 to preserve the collinear enhancement factor $\delta_{13}^2$ in the jet correlation \eqref{eq:443444}.

This dynamical ``veto'' has two consequences.

First of all, it results in {\em reduction}\/ of the hardness scale of the
parton distributions from the natural scale $Q^2$ (scale of the parton distributions in the integrated cross section)
down to the observation-induced scale $\delta_{13}^2\ll Q^2$.

Then, it introduces double logarithmic (DL) form factors of participating initial state partons,
since the transverse momentum of the jet pair can be compensated not only by a hard (energetic) parton
 from inside initial parton distributions but also by a {\em soft gluon}\/ whose radiation did not affect
 inclusive parton distributions due to real--virtual cancellation.
\smallskip

The presence of the DL form factors depending on the logarithm of a large ratio of scales,
$\ln (Q^2/\delta_{ij}^2)$, is typical for the so-called ``semi-inclusive'' processes~\cite{DDTsemi,DDT}.
\medskip

Production of massive lepton pairs in hadron collisions (the Drell--Yan process) is a classical example of a two-scale problem.
Here enter form factors of colliding quarks that depend on the ratio of
 the invariant mass $q^2$ to the transverse momentum of the lepton pair,
 $\as\ln^2(q^2/q_\perp^2)$, in the dominant kinematical region $q_\perp^2\ll q^2$:
\beeq\label{eq:DDT2}
&& \frac{d\sigma}{dq^2\, dq_\perp^2} = \frac {d\sigma_{{\mbox{\scriptsize tot}}} } {dq^2} \nonumber\\
&&\qquad   \times\frac{\partial}{\partial q_\perp^2}  \bigg\{ D_a^q \left(x_1, q_\perp^2 \right) D_b^{\bar{q}}\left(x_2, q_\perp^2 \right)
  S_q^2\left(q^2, q_\perp^2 \right)  \bigg\}\!.  \quad { }
\eeeq
$S_q$ is the double logarithmic QCD quark form factor.
\smallskip

Sudakov quark and gluon form factors can be expressed via the exponent of the total probability of the parton decay
in the range of virtualities (transverse momenta) between the two hard scales:
\begin{widetext}
\begin{subequations}\label{eq:Sudakovs}
\beeq
S_q(Q^2,\kappa^2) &=& \exp\left\{- \int_{\kappa^2}^{Q^2}\frac{dk^2}{k^2}\frac{\alpha_s(k^2)}{2\pi} \int_0^{1-k/Q} dz\, P_q^q(z) \right\}, \\
S_g(Q^2,\kappa^2) &=& \exp\left\{- \int_{\kappa^2}^{Q^2}\frac{dk^2}{k^2}\frac{\alpha_s(k^2)}{2\pi}
\int_0^{1-k/Q} dz\left[ zP_g^g(z) + n_fP_g^q(z)\right] \right\}.
\eeeq
\end{subequations}
Here $P_i^k(z)$ are the non-regularized one-loop DGLAP splitting functions (without the ``+'' prescription):

\beq\label{eq:SPLITS}
\begin{split}
  P_q^q(z) & = C_F \frac{1+z^2}{1-z},  \qquad  \quad\quad \>  P_q^g(z) =   P_q^q(1-z), \cr
  P_g^q(z) & = T_R\big[ z^2+\! (1\!-\!z)^2\big], \quad\, P_g^g(z) =  C_A\frac{1+\!z^4 +\! (1\!-\!z)^4}{z(1-z)};
\end{split}
\eeq
\end{widetext}
the upper limit of $z$-integrals properly regularizes the soft gluon
singularity, $z\to 1$
(in physical terms, it can be looked upon as a condition that the energy of a gluon should be larger than its transverse momentum, \cite{DDT}).

\bigskip

The case of hadron interactions producing large transverse momentum {\em partons}\/
(instead of colorless objects like a Drell--Yan pair or an intermediate boson) is more involved since
here the transverse momentum imbalance may be compensated by QCD radiation from the {\em final state partons}\/ too.

The azimuthal correlation between two nearly back-to-back large transverse momentum {\em particles}\/
was considered in \cite{DDT}. An analog of the ``DDT formula'' has been derived in the collinear approximation, which expression
 contained the product of four form factors, two initial parton distributions and two fragmentation functions.

\subsection{Single Logarithmic soft gluon effects}
The case when {\em jets}\/ are being reconstructed in the final state is more complicated to analyze
as it yields an answer depending on the jet finding algorithm.
The problem has been addressed by Banfi and Dasgupta in \cite{ABC} where a smart way of defining the final state jets
was formulated that permitted to write down a resummed QCD formula for soft gluon effects in ``2 partons $\to$ 2 jets'' cross sections.

Collinear logarithms due to hard splittings of the final state partons do not pose a problem: such secondary partons
 populate the jets. Partons that appear as separate out-of-jet radiation --- and are relevant
 for transverse momentum imbalance compensation --- have to be produced at sufficiently large angles with respect to the jet axis.
This is the domain of large-angle gluon radiation. Production of soft gluons in-between jets is also logarithmically enhanced
and induces single logarithmic (SL) corrections, $\left[\as\ln(Q^2/\delta^2)\right]^n$, that may also be significant and should be resummed in all orders.

Contrary to {\em collinear enhanced effects}\/ (that drive evolution of parton distributions
 and fragmentation functions and determine the Sudakov form factors), the large-angle gluon
 radiation cannot be attributed to one or another of the partons participating in the hard scattering.
It is coherent and depends on the kinematics and color topology of the hard parton ensemble as a whole. 
As a result, resummation of these SL corrections becomes a matrix problem that involves tracing various color states 
of the parton system, see \cite{ABC} and references therein.

In the present paper we concentrate on resummation of {\em collinear enhanced}\/ DL and SL terms and avoid
complications due to {\em soft}\/ SL corrections. This means ignoring color transfer effects in hard interactions.
Thus, production of four jets with large transverse momenta $j_{\perp}\sim Q$ and pair imbalances $\delta_{ij}$ will be equivalent, 
in our treatment, to production of two colorless Drell--Yan pairs with invariant masses $\cO{Q}$ and transverse momenta 
$\delta_{13}$ and $\delta_{24}$.
Generalization of the results of \cite{ABC} to the case of double parton scattering seems straightforward and should be considered separately.

\section{Generalized double parton distribution\label{Sec:EvEq}}

\subsection{Geometry of $_2$GPD}
The $_2$GPD in the expression for the multiparton production cross section has a meaning of a two body form factor when partons $1$
and $2$  receive transverse momenta $\Delta$ and $-\Delta$ leaving the hadron intact.  Nonrelativistic  analogue of this form factor
is familiar from the double scattering amplitude in the momentum space representation of the  Glauber model, see e.g.\  \cite{LevinStrikman}.
Recall  that \cite{BDFS1} the scale $\Delta$ in $_2$GPD is conjugate to  the relative transverse distance between the two partons in the
$_2$GPD in the impact parameter representation considered in \cite{TreleaniPaver,mufti,Diehl}.
\medskip

Two partons may originate from soft low-scale fluctuations inside the hadron; they can also emerge from a perturbative splitting
of a common parent parton at relatively large momentum scales.
It is clear that these two contributions to $_2$GPD will have essentially different dependence on the parameter $\Delta$.

The first contribution we will denote
\beq\label{eq:D2}
{}_{[2]}\! D^{bc}_a(x_1,x_2;  q_1^2,q_2^2; \vec{\Delta}),
\eeq
with the subscript $_{[2]}$ stressing that here the partons $b$ and $c$ emerge from the no-perturbative wave function
of the hadron $a$.
It should decrease rapidly at scales larger than a natural scale of short-range parton correlation in a hadron
(this scale may be slightly different for quarks and gluons and could in principle be significantly larger than the
$1/r_N$ as there exists another non-perturbative scale of the chiral symmetry breaking which maybe as large as 700 MeV).

$_2$GPD should rapidly decrease
for $\Delta^2 \ge 1.5/\left<r_{t \, g}^2\right>$ where $\left<r_{t \, g}^2\right>^{1/2}$ is the transverse gluonic radius of the nucleon.
In the mean field approximation when the correlations are neglected,
it can be approximated by a factorized expression  \cite{BDFS1}
\begin{subequations}\label{eq:2DFdef}
\beeq\label{eq:2Ddef}
 {}_{[2]}\! D^{b,c}_a(x_1,x_2;  q_1^2,q_2^2; \vec{\Delta}) &=& F_g(\Delta^2; x_1,q_1^2)F_g(\Delta^2; x_2,q_2^2)
\nonumber\\
& \times&  G_a^b(x_1,q_1^2) G_a^c(x_2,q_2^2) ,
\eeeq
where $G$ are the single-parton distributions and the two-gluon form factor $F$
can be parametrized as
\beq\label{eq:Fdef}
   F_g(\Delta^2) \>=\> \left(1+\frac{\Delta^2}{m^2_g(x)}\right)^{-2}.
\eeq
\end{subequations}
The parameter $m_g^2 $ is of the order of 1 GeV$^2$ for $x \sim 10^{-2}$ and gradually drops with decrease of $x$
and with increase of virtuality \cite{FSW}.

\medskip

The second contribution we will denote
\beq\label{eq:D1}
{}_{[1]}\! D^{b,c}_a(x_1,x_2;  q_1^2,q_2^2; \vec{\Delta}),
\eeq
where the subscript  $_{[1]}$ stands as a reminder of the fact that $b$ and $c$ originate from
 perturbative splitting of a single parton from the hadron wave function.
This contribution is practically $\Delta$-independent and should decrease with $\Delta$ much more slowly, due to logarithmic pQCD effects.
(A steep power falloff starts only when $\Delta^2$ exceeds the relevant hard scale, $q^2$.)

\subsection{On the geometrical enhancement of the interference effects due to pQCD correlations}

By total cross section in the present context we mean the  back-to-back 4-jet cross section integrated
over pair jet imbalances in the dominant logarithmic region $\delta_{ik}\ll j_{i\perp} \simeq j_{k\perp}$.

We start by noting that the product of two small-distance parton fluctuations, ${}_{[1]} \!D \times {}_{[1]}\! D $,
does not contribute to the process we are interested in.
Indeed, in this case the integral over $\Delta$ in Eq.~\eqref{eq:1} formally diverges and yields a hard scale (instead of $R^{-2}$) in the numerator. 
This means a significant contribution to the cross section but not the one we are looking for.
Below in Section~\ref{Sec24} we will explicitly verify that a double hard collision of two parton pairs each of which originates 
from perturbative splitting, lacks the back-to-back enhancement.
In fact, the product ${}_{[1]}\! D \times {}_{[1]}\!  D$ corresponds to a one-loop correction to the ``leading twist''
perturbative production of four jets in a hard collision of two partons (``two-to-four'') whose distribution
is {\em smooth}\/ in the back-to-back region and as such {\em gets subtracted}\/ as  background.

Keeping this in mind, the back-to-back 4-jet production cross section is proportional to the inverse ``interactions area'' $S$
described by the expression
\begin{widetext}
\beq
\frac{1}{S}=\int \frac{d^2\Delta}{(2\pi)^2} \biggl(  {}_{[2]}\!D_{a}(\Delta)\,  _{[2]}\!D_{b}(\Delta) \>+\>
_{[2]}\!D_{a}(\Delta)\,  _{[1]}\!D_{b}(\Delta)  \>+\>
_{[1]}\!D_{a}(\Delta)\,  _{[2]}\!D_{b}(\Delta)  \biggr),
\label{s}
\eeq
\end{widetext}
where indices $a$, $b$ mark two interacting nucleons. 
This expression is somewhat symbolic; a careful analysis of the ``interaction area'' 
will be carried out below in Sec.~\ref{Sec:4324} (see Eq.~\eqref{stot}).

The first term in Eq.~\eqref{s} we will refer to as a ``four-to-four'' process: two partons from the wave functions 
of the hadron $a$ interact with two partons from the hadron $b$ producing four jets.
The second and the third terms in Eq.~\eqref{s} describe hard collisions of one parton from one hadron 
with two partons from the second hadron. 
Until recently, these ``three-to-four'' processes were commonly ignored in the literature 
(see, however,  \cite{Ryskin}
). 
At the same time, they turn out to be somewhat enhanced.

Indeed, the contribution due to four-to-four processes to the geometrical factor Eq.~\eqref{s} is given by
\begin{subequations}
\beq \label{eq:44est}
   \int \frac{d^2\Delta}{(2\pi)^2}\> F_g^2(\Delta^2)\times F^2_g(\Delta^2)=\frac{m^2_g}{7\pi}.
\eeq
Fast decrease of the product of two squared form factors leads to fast convergence of the integral whose
median is positioned at a value as low as $\Delta^2 \approx 0.1 m_g^2$.

\medskip

The case of the three-to-four process is different.
This process corresponds, as we explained above, to interaction of the offspring of the perturbative splitting 
of a parton from the wave function of one hadron, with two partons from the non-perturbative wave function 
of the second colliding hadron.
On the side of $_{[1]}D$, the parameter $\Delta$ enters ``perturbative loop'' due to parton splitting 
and as a result the dependence of $_{[1]}D$ on $\Delta$ turns out to be only logarithmic,
that is, parametrically much slower than that of the non-perturbative form factor $F_g(\Delta^2)$.
Thus, the three-to-four contribution to the double interaction cross section reduces to
\beq
   \int\!\! \frac{d^2\Delta}{(2\pi)^2} \,{} {}_{[1]}D(x_1,x_2;\Delta)\, F^2_g(\Delta^2)
  \! \simeq  \left.{} _{[1]}D\right|_{\Delta\!=\!0} \!\int\!\! \frac{d^2\Delta}{(2\pi)^2}F^2_g(\Delta^2) ,
\eeq
(where we have neglected the logarithmic $\Delta$-dependence of $_{[1]}D$).
This corresponds to the fact that in the impact parameter space, the distance between partons coming 
from a perturbative splitting is much smaller than the hadron size, so that the answer is proportional 
to the density of non-perturbative two-parton correlation at small distances --- ``in the origin'':
\beq  \label{eq:34est}
  \int \frac{d^2\Delta}{(2\pi)^2}F^2_g(\Delta^2) \>=\> \frac{m^2_g}{3\pi}.
\eeq
\end{subequations}
Comparison with the estimate \eqref{eq:44est} shows that the contribution to the cross section 
of the ``interference term'' $1+2\to 4$ is enhanced, relative to the $2+2\to4$ process,
by the factor
\beq
   \frac{7}{3}\times 2\sim 5.
\eeq
(For the case of the Gaussian form factors this enhancement is 15\%\ smaller --- a factor of 4.)
This estimate was obtained for the case when all four partons participating in the hard collisions are gluons.
A detailed numerical study of the $x_i$-dependence of
the effective interaction area $S$
will be presented in the paper under preparation.

So we conclude that the three-to-four processes may provide a sizable contribution to the cross section
even if they constitute a small correction to $_2$GPD.


\subsection{Perturbative QCD effects in $_2$GPD}

Thus, we represent the generalized double parton distribution $_2$GPD as a sum of two terms:
\beeq\label{eq:2terms}
D^{b,c}_a(x_1,x_2;  q_1^2,q_2^2; \vec{\Delta}) &=&  {}_{[2]} \!D^{b,c}_a(x_1,x_2;  q_1^2,q_2^2; \vec{\Delta}) \nonumber\\
&+&  {}_{[1]} \!D^{b,c}_a(x_1,x_2;  q_1^2,q_2^2; \vec{\Delta}) . \quad{ }
\eeeq

The term ${}_{[2]} D$ describes the distribution of two partons from the non-perturbative wave function of the
 hadron $a$ that are {\em independently evolved}\/  to large perturbative scales $q_1^2$ and $q_2^2$ according to the
 standard one-parton evolution equation.
The perturbative evolution involves momentum scales much larger than the hadron wave function correlation scale
$\lrang{\Delta^2}\sim Q_0^2$. Therefore, the evolution practically does not affect the $\Delta$-dependence of the two-parton spectrum.
This piece of the $_2$GPD acquires but a mild additional logarithmic dependence {\em at the tail}\/ of
 the $\Delta$-distribution in addition to a non-perturbative power falloff \eqref{eq:2DFdef}.

The integral QCD evolution equation for ${}_{[2]} D^{bc}_a$ reads
\begin{widetext}
\beq\label{eq:termNP}
\begin{split}
{}_{[2]}\! D^{b,c}_a( \, & x_1, x_2; q_1^2,q_2^2; \vec{\Delta})  \>=\> S_b(q_1^2, Q_{\min}^2)S_c(q_2^2, Q_{\min}^2)\,
{}_{[2]}\!D^{b,c}_a(x_1,x_2;  Q_0^2,Q_0^2; \vec{\Delta}) \cr
& + \sum_{b'} \int_{Q_{\min}^2}^{q_1^2} \frac{dk^2}{k^2}\frac{\as(k^2)}{2\pi} \, S_b(q_1^2,k^2)
  \int\!\frac{dz}{z}\>P_{b'}^{b}\!\!\left(z\right)  {}_{[2]}\!D^{b',c}_{a}\left(\frac{x_1}{z},x_2;k^2,q_2^2; \vec{\Delta}\right) \cr
& + \sum_{c'} \int_{Q_{\min}^2}^{q_2^2} \frac{dk^2}{k^2}\frac{\as(k^2)}{2\pi} \, S_c(q_2^2,k^2)
  \int\!\frac{dz}{z}\>P_{c'}^{c}\!\!\left(z\right)  {}_{[2]}\!D^{b,c'}_{a}\left(x_1,\frac{x_2}{z};q_1^2,k^2; \vec{\Delta}\right) .
\end{split}
\eeq
\end{widetext}
Here $P_i^k(z)$ are the non-regularized one-loop DGLAP splitting functions \eqref{eq:SPLITS}
and $S_i$ --- the double logarithmic Sudakov parton form factors  defined in Eq.~\eqref{eq:Sudakovs}.

The lower limit of the perturbative evolution in Eq.~\eqref{eq:termNP},
\beq\label{eq:Qmindef}
    Q_{\min}^2 = \max(Q_0^2, \Delta^2) \>\simeq\> Q_0^2+\Delta^2 ,
\eeq
is the only source of additional (logarithmic) $\Delta$-dependence.
It emerges when $\Delta^2$ exceeds --- and substitutes --- the starting non-perturbative scale $Q_0^2$ of the perturbative evolution.

\bigskip
The second term in Eq.~\eqref{eq:2terms},  ${}_{[1]} D^{bc}_a$, represents the small-distance correlation between the two partons
that emerge from a perturbative splitting of a common parent parton taken from the hadron wave function.
It can be expressed in terms of standard inclusive single-parton distributions as follows
\begin{widetext}
\beq\label{eq:termPT}
\begin{split}
{}_{[1]} \!D^{b,c}_a(x_1,x_2;   q_1^2,q_2^2; \vec{\Delta}) &= \sum_{a',b',c'} \int_{Q_{\min}^2}^{\min{(q_1^2,q_2^2)}}\frac{dk^2}{k^2}\frac{\as(k^2)}{2\pi}  \int\!\frac{dy}{y^2}  \>G^{a'}_a(y;k^2 ,Q_0^2) \cr
& \times  \int\!\frac{dz}{z(1-z)}
 \> P_{a'}^{b'[c']}\!\!\left(z\right) \> G^{b}_{b'}\left(\frac{x_1}{zy};q_1^2,k^2\right) G^{c}_{c'}\left(\frac{x_2}{(1-z)y};q_2^2,k^2\right).
\end{split}
\eeq
\end{widetext}
The $\Delta$-dependence of ${}_{[1]} D$ is very mild as it emerges solely from the lower limit of the logarithmic transverse momentum integration $Q^2_{\min{}}$.

\section{Analysis of perturbative two-parton correlations\label{Sec:PT}}

\subsection{$3\to4$\label{SubSec:PT34}}
\begin{figure}[h]
\begin{center}
 \includegraphics[height=5.5cm]{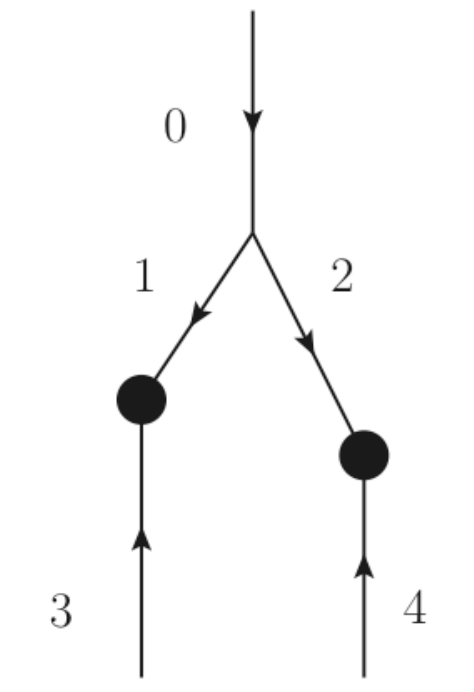}
\end{center}
\caption{\label{Fig34} $3\to4$}
\end{figure}

Let us analyze the lowest order interaction amplitude shown in Fig.~\ref{Fig34} that produces a double hard collision and involves parton splitting.

We express parton momenta $k_i$ in terms of the Sudakov decomposition using the light-like vectors $p_a$, $p_b$ along the incident hadron momenta:
\begin{subequations}
\beeq
k_1 &=& x_1p_a + \beta p_b + k_{\perp}, \quad  k_3 \simeq (x_3-\beta)p_b; \nonumber \\
k_2 &=& x_2p_a - \beta p_b - k_{\perp}, \quad k_4 \simeq (x_4+\beta)p_b; \nonumber \\
 \vec{k}_\perp &=& \vec{\delta}_{12} = -\vec{\delta}_{34} \> (\delta'\equiv0); \> k_0 \simeq(x_1+x_2)p_a. \nonumber
\eeeq
\end{subequations}
Here $k_0$, $k_3$ and $k_4$ are momenta of incoming (real) partons, and $k_1$ and $k_2$ --- virtual ones.
Light-cone fractions $x_i$, i=1,..4, are determined by jet kinematics (invariant masses and rapidities of jet pair). 
The fraction $\beta$ that measures the {\em difference}\/ in longitudinal momenta of the two partons coming from the hadron $b$, is arbitrary.
Fixed values of the parton momenta $x_3-\beta$ and $x_4+\beta$ correspond to the plane wave description of the scattering process in which the longitudinal distance between the two scatterings is arbitrary. This description does not correspond to the physical picture of the process we are interested in. 
In order to ensure than the partons $3$ and $4$ originate form the  {\em same hadron}\/ of finite size, we have to introduce an integration over 
$\beta$ in the {\em amplitude}, in the  region $\beta=\cO{1}$.
\medskip

The Feynman amplitude contains the product of two virtual propagators. The virtualities $k_1^2$ and the $k_2^2$ that enter the denominator 
of the amplitude in terms of the Sudakov variables read
\beq
  k_1^2 = x_1\beta s -k_\perp^2, \>\>  k_2^2 = -x_2\beta s -k_\perp^2, \nonumber
\eeq
with $s=2p_ap_b$ and $k_\perp^2\equiv (\vec{k}_{\perp})^2>0$ the square of the two-dimensional transverse momentum vector.

A singular contribution we are looking for originates from the region $\beta\ll1$, so that precise form of the longitudinal smearing 
does not matter and the integral yields
\beq
N\!\! \int\! \frac{d\beta}{(x_1\beta s -k_\perp^2 +\!i\epsilon)(-x_2\beta s -k_\perp^2 +\!i\epsilon)}
=  \frac{2\pi i N}{(x_1\!+\!x_2)}\frac{1}{k_\perp^2}. \nonumber
\eeq
The numerator of the amplitude is proportional to the {\em first power of the transverse momentum}\/ $k_\perp$.
As a result, the squared amplitude (and thus the differential cross section) acquires the necessary factor $1/\delta^2$ 
that enhances the back-to-back jet production.

\subsection{$2\to4$\label{Sec24}}
\begin{center}
\begin{figure}[h]
 \includegraphics[height=5.5cm]{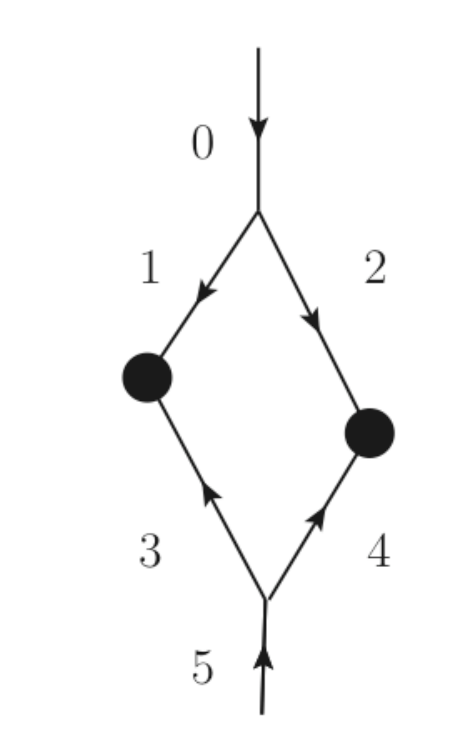}
\caption{\label{Fig24} $2\to4$}
\end{figure}
\end{center}

Now we should verify that the diagram of Fig.~\ref{Fig24} where both incident partons split, and their offspring engage
into double hard scattering, does not favor back-to-back jet kinematics. In other words, it does not lead
to a small imbalance factor $1/\delta^2$ in the differential cross section.

\noindent
Sudakov decomposition:
\[
\begin{split}
& \!\!\! k_1= (x_1 \! -\! \alpha)p_a + \beta p_b + k'_{\perp}, \, k_3 = (x_3\! -\! \beta)p_b + \alpha p_a  - k_\perp;  \cr
& \!\!\!  k_2 = (x_2\! +\! \alpha)p_a - \beta p_b - k'_{\perp},\, k_4 = (x_4\! +\! \beta)p_b -\alpha p_a  +k_\perp;  \cr
& \!\!\!  \vec{k}'_\perp - \vec{k}_\perp = \vec{\delta}_{12} = -\vec{\delta}_{34} \quad (\delta'\equiv0);  \cr
& \!\!\!  k_0  \simeq (x_1 + x_2)p_a,\, k_5\simeq (x_3 + x_4)p_b.
\end{split}
\]
\smallskip

\noindent
This is a loop diagram and it contains explicit integration over the loop momentum:
\[
s\int\frac{d\alpha\, d\beta}{(2\pi)^2i}\int\frac{d^2k_\perp d^2k'_\perp}{(2\pi)^2}\delta^2(\vec{k}_\perp +\vec{\delta} - \vec{k}'_\perp) .
\]
To get an enhanced contribution we have to have parton virtualities that enter the denominator of the Feynman amplitude to be relatively small, 
of the order of ${\delta^2}\ll Q^2$.
This implies $|\alpha|, |\beta| \ll 1$ in the essential integration region. Adopting this approximation, we can simplify parton propagators 
and reduce the longitudinal momentum integrations to the product of two independent integrals:
\[
\begin{split}
\frac{i}{s}  & \int\frac{d\beta}{2\pi i} \frac{s}{(\beta x_1s \!-\! k_\perp^2 + i\epsilon)(\beta x_2s \!+\! k_\perp^2 - i\epsilon)} \cr
 \times & \int\frac{d\alpha}{2\pi i}   \frac{s}{(\alpha x_3s \!-\! k'^2_\perp + i\epsilon)(\alpha x_4s \!+\! k'^2_\perp - i\epsilon)} \cr
& = \frac{i}{(x_1+x_2)(x_3+x_4)s}\>\frac{1}{k_\perp^2\, k'^2_\perp}.
\end{split}
\]
The remaining transverse momentum integration takes the form
\beq\label{eq:kperpint}
\int\frac{d^2k_\perp}{(2\pi)^2} \> \frac{V}{{\vec{k}_\perp}^2(\vec{k}_\perp + \vec\delta)^2}
\eeq
Due to gauge invariance the numerator of the diagram --- the ``vertex factor'' $V$ --- is linear in transverse momenta of the loop partons: 
$V\propto k_\perp^\mu k'^\nu_\perp$. Therefore, the integral \eqref{eq:kperpint} produces no more than a logarithmic enhancement factor, 
$\ln(Q^2/\delta^2)$, instead of the power back-to-back singularity $Q^2/\delta^2$ we were looking for.

So, the diagram Fig.~\ref{Fig24} with double parton splitting constitutes but a negligible loop correction
to the usual ``hedgehog'' 4-jet kinematics typical for $2\to 4$ QCD processes.
The fact that this loop diagram does not produce a pole singularity in $\delta^2$ could have been extracted, 
e.g., from numerical studies of double $Z$-boson production in two-parton collisions \cite{GloverBij} and, more generally, 
of multi-leg parton amplitudes \cite{NagySoper}. 

The logarithmic character of this correction has been recently confirmed by the systematic study of ``box integrals'' in \cite{stirling1}.

The presence of the double parton splitting contribution of Fig.~\ref{Fig24} is being treated in the literature as a source 
of potential problem of double counting (see, e.g., \cite{NagySoper,CaSaSa,DiehlSchafer}). The present paper solves this problem.   

\subsection{$3\to4$ with additional parton emission \label{subsection34-add}}
We have to return now to the $3\to4$ process and examine the possibility of producing an additional parton,
in collinear enhanced manner, in order to lift off the Born level kinematical constraint $\delta'=0$.

\begin{widetext}
\begin{center}
\begin{figure}[h]
 \includegraphics[height=6cm]{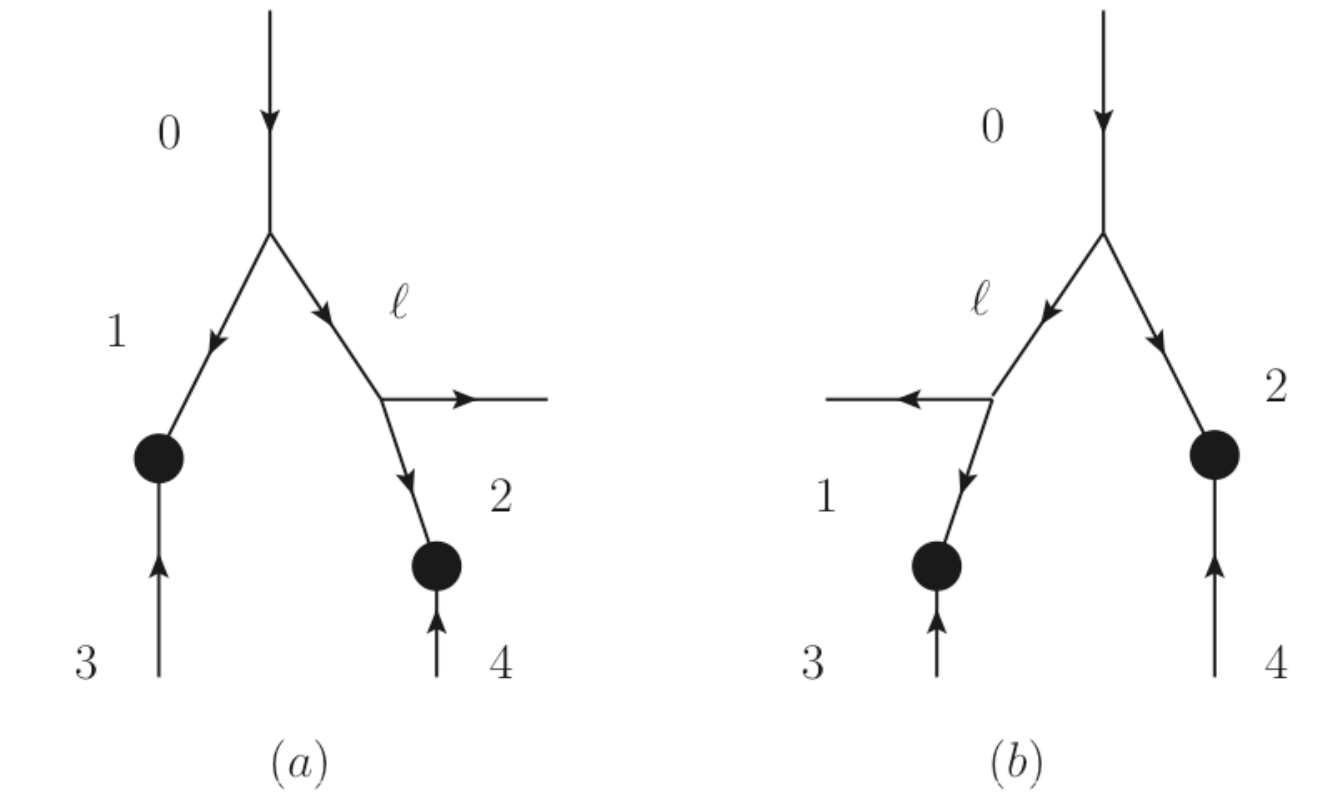}
\caption{\label{Fig34-1}Three-to-four amplitudes with extra emission from inside the splitting fork}
\end{figure}
\end{center}
\end{widetext}

Consider the diagram of  Fig.~\ref{Fig34-1}a.
The momenta of quasi-real colliding partons are
\[
  k_0 \simeq (x_1+x_2+\alpha)p_a;  \>  k_3 \simeq (x_3-\beta)p_b, \>  k_4 \simeq (x_4+\beta + \frac{\delta'^2}{\alpha s})p_b,
\]
and the radiated on-mass-shell parton carries momentum
\[
    \ell - k_2 \>=\> \alpha p_a + \frac{\delta'^2}{\alpha s}p_b - \delta', \qquad \delta' = \delta_{13}+\delta_{24}.
\]
We have three virtual propagators subject to integration over $\beta$:
\begin{subequations}
\beeq
k_1 &=& x_1p_a + \beta p_b + \delta_{13}, \\
k_2 &=& x_2p_a - \left(\beta + \frac{\delta'^2}{\alpha s}\right)p_b + \delta_{24}, \\
\ell &=& (x_2+\alpha) p_a   -\beta p_b   - \delta_{13}.
\eeeq
\end{subequations}
Closing the contour around the pole $k_1^2+ i\epsilon=0$, we obtain $\beta s= \delta_{13}^2/x_1$ and
\begin{subequations}\label{eq:34-virts}
\beeq
\label{eq:34-virts-1}
-\ell^2 \!\!&=&\!\! (x_2+\alpha)\beta s + \delta_{13}^2 = \delta_{13}^2 \cdot \frac{x_1+x_2+\alpha}{x_1}  , \\
\label{eq:34-virts-2}
-k_2^2 \!\!&=&\!\!   x_2\!\!\left(\!\beta \!+\! \frac{\delta'^2}{\alpha s} \!\right)\!s +  \delta_{24}^2
\!=\!\! \frac{x_1\delta_{24}^2\!+\! x_2\delta_{13}^2}{x_1} +\! \frac{x_2}{\alpha}\delta'^2 \!\!. \quad{ }
\eeeq
\end{subequations}
Taken together with the residue of the $\beta$-integration, $1/x_1$, Eq.~\eqref{eq:34-virts-2} produces 
the universal factor present in all the amplitudes considered, (including the diagrams with parton emission off the external lines, 
see Fig.~\ref{Fig34-2} below):
\beq\label{eq:Pdef}
   P^{-1} = x_1\delta_{24}^2 + x_2\delta_{13}^2 + \frac{x_1x_2}{\alpha}\delta'^2\,.
\eeq
We observe that both propagators \eqref{eq:34-virts} are enhanced in the back-to-back kinematics.
The amplitude of Fig.~\ref{Fig34-1}a gives a double collinear enhanced contribution to the cross section in the region
\begin{subequations}\label{eq:34-regions}
\beq\label{eq:34-region1}
   \delta_{13}^2\> \ll\> \delta_{24}^2\,\simeq\, \delta'^2.
\eeq
The inequality \eqref{eq:34-region1} corresponds to the following physical picture.
An incident parton $k_0$ splits into $k_1$ and $\ell$ {\em early}, at time $\cO{\sqrt{s}/\delta_{13}^2}$ corresponding to some 
comparatively low perturbative scale $\delta_{13}$.
At this time scale the parton $1$ collides with $3$, while the parton $\ell$ keeps evolving and scatters off $4$ with 
a much larger momentum transfer $\delta_{24}$. Evolution of the parton $\ell$ in between these two scales is the origin 
of probable (logarithmically enhanced) production of additional parton(s).

Analogously, the diagram Fig.~\ref{Fig34-1}b  with a parton produced off the virtual line $1$ contributes in the complementary kinematical region
\beq\label{eq:34-region2}
   \delta_{24}^2\> \ll\> \delta_{13}^2\,\simeq\, \delta'^2.
\eeq
\end{subequations}
Full perturbative analysis of the production of a parton from inside the ``splitting fork'', Fig.~\ref{Fig34-1}, 
together with emission off the incoming line ``0'' (to be treated below in Sec.~\ref{Sec:324}) is sketched in the Appendix.


\section{Differential distribution\label{Sec:4324}}

\subsection{$2+2\to4$\label{Sec:424}}
Now that we know the structure of the perturbative corrections to $_2$GPD, Eqs.~\eqref{eq:2terms}--\eqref{eq:termPT},
we are in a position to write down the generalization of the DDT formula \eqref{eq:DDT2} for (the first contribution to)
the differential cross section of 4-jet production in nearly back-to-back kinematics.
It reads
\begin{widetext}
\beeq\label{eq:DDT4}
\pi^2\frac{d\sigma^{(4\to4)}}{d^2\delta_{13}\, d^2\delta_{24}} &=&
 \frac {d\sigma_{{\mbox{\scriptsize part}}}
 } {d\hat{t}_1\,d\hat{t}_2}
\cdot \frac{\partial}{\partial\delta_{13}^2}
 \frac{\partial}{\partial\delta_{24}^2} \bigg\{
{}_{[2]}\!D_a^{1,2}(x_1,x_2;\delta_{13}^2, \delta_{24}^2 )  \times {}_{[2]}\!D_b^{3,4}(x_3,x_4;\delta_{13}^2, \delta_{24}^2 )
\nonumber\\
&\times&  S_1\left({Q^2},\delta_{13}^2 \right)S_3\left({Q^2},{\delta_{13}^2}\right)
 \times S_2\left({Q^2},{\delta_{24}^2}\right) S_4\left({Q^2},\delta_{24}^2\right)   \bigg\} .
\eeeq
\end{widetext}
Here $d\sigma_{{\mbox{\scriptsize part}}}$ is the cross section of double hard parton scattering, and $S_i$ stand for 
Sudakov form factors of four participating partons. Sum over parton species and convolution over $\vec{\Delta}$ as in Eq.~\eqref{eq:1} is implied.

Taking derivative over the scale $\delta^2$ of the function depending on $\as\log\delta^2$, produces the factor $\as/\delta^2$. 
Differentiating the Sudakov form factor of a given parton describes the situation when the jet imbalance is compensated by radiation 
of a {\em soft gluon}\/ off this parton.  Differentiation of the parton distribution corresponds to the situation when a {\em hard parton}\/ takes the recoil.

\subsection{$1+2\to4$\label{Sec:324}}
The differential transverse momentum imbalance distribution due to the cross-terms ${}_{[2]}D \times {}_{[1]}D$ contains two pieces.
\subsubsection{Two compensating partons}
The first one has the same structure as Eq.~\eqref{eq:DDT4}:
\begin{widetext}
\beeq\label{eq:DDT31}
\pi^2\frac{d\sigma^{(3\to4)}_1}{d^2\delta_{13}\, d^2\delta_{24}} &=&
 \frac {d\sigma_{{\mbox{\scriptsize part}}}
 } {d\hat{t}_1\,d\hat{t}_2}
\cdot \frac{\partial}{\partial\delta_{13}^2}
 \frac{\partial}{\partial\delta_{24}^2} \bigg\{
{}_{[1]}\!D_a^{1,2}(x_1,x_2;\delta_{13}^2, \delta_{24}^2 ) \cdot {}_{[2]}\!D_b^{3,4}(x_3,x_4;\delta_{13}^2, \delta_{24}^2 )
\nonumber\\
&\times&  S_1\left({Q^2},\delta_{13}^2 \right)S_3\left({Q^2},{\delta_{13}^2}\right)
 \cdot S_2\left({Q^2},{\delta_{24}^2}\right) S_4\left({Q^2},\delta_{24}^2\right)   \bigg\} .
\eeeq
\end{widetext}
Sum over parton species and convolution over $\vec{\Delta}$  is implied as above in Eq.~\eqref{eq:DDT4}.

As above, taking the derivatives in Eq.~\eqref{eq:DDT31} corresponds to fixing transverse momenta 
of two final state partons that compensate jet pair imbalances:  $-\vec{\delta}_{13}$ and $-\vec{\delta}_{24}$. 

Consider now the correlation term $_{[1]}D$ of Eq.~\eqref{eq:DDT31}. 
If we apply the derivatives to the parton distributions $D$ in the {\em integrand}\/ of Eq.~\eqref{eq:termPT} for the correlation term $_{[1]}D$, 
this contribution will correspond to production of two momentum compensating quanta in the course of evolution of the system of two partons 
with account of {\em small-distance perturbative correlation}\/ between them.
In this case the scale of the core parton splitting stays smaller than the two external scales ${\delta}_{13}$, ${\delta}_{24}$, and is being integrated over. 
 
\subsubsection{One compensating parton}
The correlation term \eqref{eq:DDT31} contains an additional option. 
Namely, instead of creating intermediate state partons that keep evolving up to external scales, 
the perturbative splitting may produce that very parton that gets engaged in the hard scattering.
This possibility is also contained in Eq.~\eqref{eq:DDT31}: it corresponds to the differentiation of the {\em upper limit}\/
of the virtuality integral in Eq.~\eqref{eq:termPT} over the {\em smaller}\/ of the two imbalances, $\min\{\delta_{13}^2, \delta_{24}^2\}$.
One of the two parton distribution functions in the integrand then collapses to $\delta(1-x/y)$.
Taking the second derivative over the {\em larger}\/ imbalance of the second parton distribution produces 
the contribution described by the diagram of Fig.~\ref{Fig34-1} that we have discussed above 
in Section~\ref{subsection34-add}.

\subsection{$\bf 1+2\to4$, endpoint contribution\label{Sec:324end}}

Finally, there is a possibility that {\em both}\/ partons emerging from the perturbative splitting in $_{[1]}D$ 
experience hard collisions straight away, without any further evolution. 
Jet imbalances stemming from such an eventuality are no longer independent but, on the contrary, 
are strongly correlated. 
At the ``Born level'' one has
\[
\frac{d\sigma}{d^2\delta_{13}d^2\delta_{24}} \propto \frac{\as}{\delta^2}\delta(\vec{\delta}_{13}+\vec{\delta}_{24}),
\quad \delta^2 \equiv  \delta_{13}^2= \delta_{24}^2.
\]
With account of additional radiation, the delta-function gets replaced by the pole enhancement of 
the differential cross section in the {\em imbalance of imbalances}\/ in a specific region of jet momenta 
when the pair imbalances are practically equal-and-opposite::
\[
\frac{d\sigma}{d^2\delta_{13}d^2\delta_{24}} \propto \frac{\as^2}{\delta^2\,\delta'^2}, \quad 
\delta'^2\ll \delta^2 \equiv  \delta_{13}^2\simeq \delta_{24}^2.
\]
Importantly, this contribution is also double-collinear enhanced and therefore has to be taken into full consideration. 
This eventuality is not incorporated in Eq.~\eqref{eq:DDT31} and should be taken care of separately. 

In this kinematical region the parton that carries the compensating transverse momentum 
$-\vec{\delta}'$ can be produced off one of the external legs, as shown in Fig.~\ref{Fig34-2}.
\begin{widetext}
\begin{center}
\begin{figure}[h]
 \includegraphics[height=6cm]{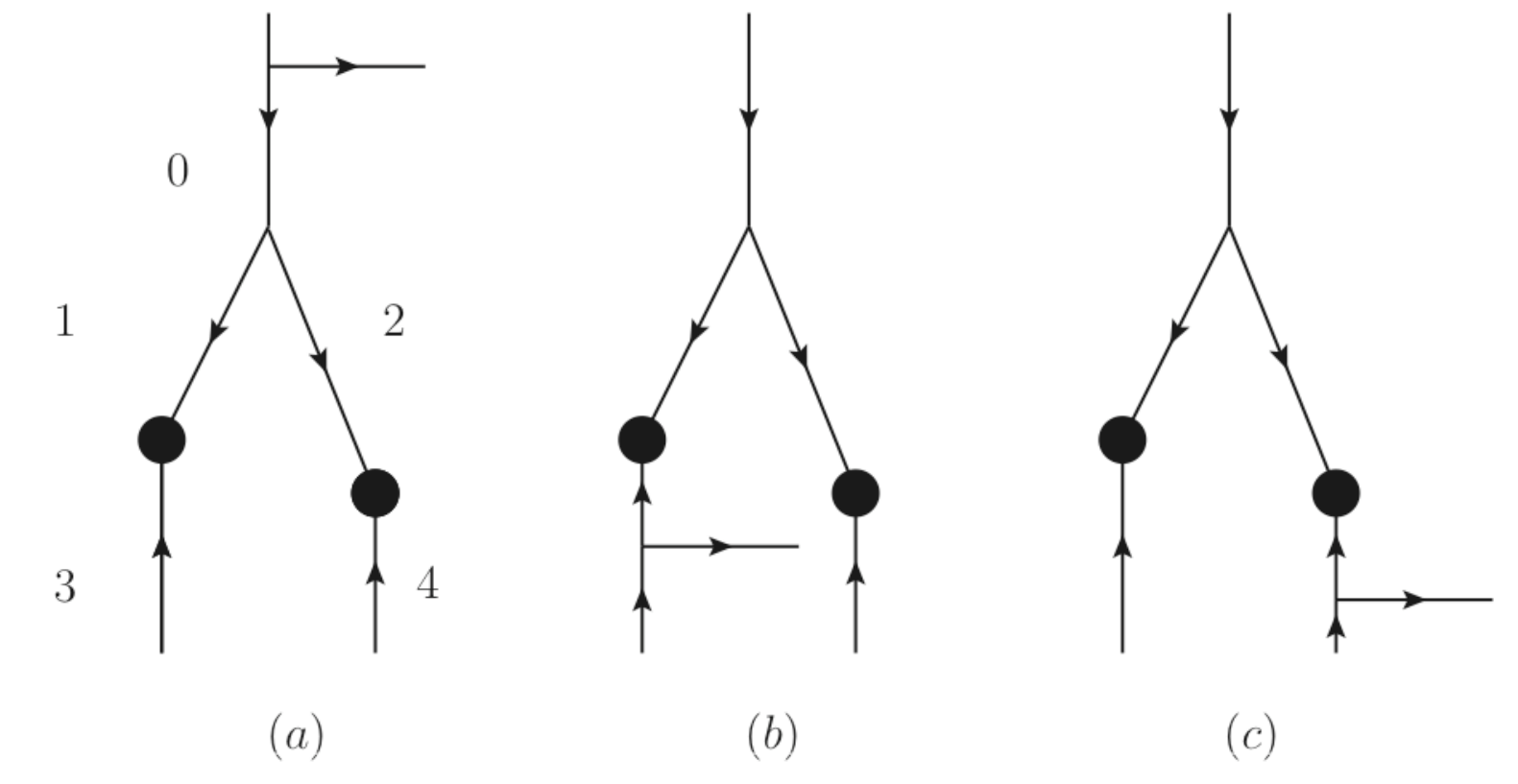}
\caption{\label{Fig34-2} $3\to4$ with real parton emission off the external lines}
\end{figure}
\end{center}

The corresponding contributions to the differential 4-jet production cross section
can be combined into the following relatively compact expression:
\beq\label{eq:DDT32}
\begin{split}
 \frac{\pi^2\> d\sigma^{(3\to4)}_2}{d^2\delta_{13}\, d^2\delta_{24}} \>&=\>\>
 \frac {d\sigma_{{\mbox{\scriptsize part}}}
 } {d\hat{t}_1\,d\hat{t}_2}\> \cdot \>   \frac{\as(\delta^2)}{2\pi\, \delta^2} \,  \sum_c  P_{c}^{1,2}\!\!
    \left(\frac{x_1}{x_1+x_2}\right) S_1(Q^2,\delta^2)\, S_2(Q^2,\delta^2) \cr
&\times \frac{\partial}{\partial\delta'^2}
 \bigg\{ S_c(\delta^2\!,\delta'^2) \frac{G_a^{c}(x_1\!+\! x_2;\delta'^2\! ,Q_0^2)}{x_1+x_2}
 S_3(Q^2\! ,\delta'^2) S_4(Q^2\! ,\delta'^2)  \times \!{}_{[2]}\!D_b^{3,4}(x_3,x_4;\delta'^2\! , \delta'^2 ) \bigg\}
\> +  \> \big\{ a \leftrightarrow b \big\}.
\end{split}
\eeq
\end{widetext}
Once again, integral over $\vec\Delta$ on the r.h.s.\ is implicit here.
In this formula the splitting of the parton ``c'' from the hadron ``a'' is written explicitly, represented by the splitting function $P_c^{1,2}$.
The splitting occurs at the virtuality scale $\delta^2$.
However, accompanying production of secondary real particles is vetoed starting from a much smaller scale, $\delta'^2\ll \delta^2$.
This explains appearance of the Sudakov form factors of five participating partons in \eqref{eq:DDT32}.

In the {\em impulse approximation}, there is one parton produced with the transverse momentum $-\vec{\delta}'$ compensating
the 4-jet imbalance, while transverse momenta of all other real partons are smaller: $k_{\perp i}^2\ll \delta'^2$.
Production probability of this parton integrated over its energy (rapidity) is embedded into the derivative of the product
of three Sudakov form factors and of the initial parton distributions depending on $\delta'^2$ as the upper evolution scale.
As before in Eq.~\eqref{eq:DDT4}, by differentiating form factors we obtain soft gluon radiation off three external lines ``c" (``0''), ``3'' and ``4''.
Differentiation of the parton distribution functions describes real parton production due to ``hard'' splittings.

The full answer is given by the sum of Eqs.~\eqref{eq:DDT4}--\eqref{eq:DDT32}:
\beq\label{eq:sumsigmas}
 \frac{\pi^2\> d\sigma^{(4\to4)}}{d^2\delta_{13}\, d^2\delta_{24}}
 +  \frac{\pi^2\> d\sigma^{(3\to4)}_1}{d^2\delta_{13}\, d^2\delta_{24}}
 + \frac{\pi^2\> d\sigma^{(3\to4)}_2}{d^2\delta_{13}\, d^2\delta_{24}} .
\eeq

\subsection{Distribution over jet imbalances in the extreme back-to-back limit $\delta^2, \delta'^2 \to 0$  \label{Sec:flatten}}

As we have repeatedly stressed above, double parton collisions constitute but a small correction to 4-jet production
if characterized by their contribution to the total cross section.
However, ``higher twist''  four-to-four and three-to-four processes are specific in that they populate 
the kinematical region of relatively small pair jet imbalances, 
$\delta_{13}^2, \delta_{24}^2, \delta'^2=(\vec\delta_{13}+\vec\delta_{24})^2 \ll Q^2$.

Being a multi-scale hard process, back-to-back jet production cross section possesses double logarithmic parton form factors.
The Sudakov suppression is the price one pays for having the differential distribution peaked, $d\sigma\propto \delta^{-2}$, 
which enhancement implies vetoing the bremsstrahlung radiation of gluons with transverse momenta in a broad interval ranging 
from $k_\perp^2> \delta^2$ all the way up to $k_\perp^2<Q^2$.

The formulas \eqref{eq:DDT4}--\eqref{eq:DDT32} for the differential imbalance spectra were derived in the impulse approximation
in which it is a single parton (soft gluon) that compensates the transverse momentum imbalance, while radiation of other partons 
with $k_\perp^2> \delta^2$ is strictly vetoed.
However, in the extreme $\delta^2\to0$ limit the form factor suppression becomes so strong as to override the pole enhancement.
In these conditions the impulse approximation is no longer valid, and instead of {\em vetoing}\/ accompanying radiation one has 
to impose the condition that the real partons carry, in aggregate, the given transverse momentum: $\sum \vec{k}_{\perp i}=-\vec\delta $.
This implies exponentiating soft gluon radiation in the {\em impact parameter space}, and evaluating the inverse Fourier transform 
to access the resulting $\delta$-spectrum \cite{ParPet,pt0,CoSoSt}. The result is the {\em flattening}\/ of the spectrum in the $\delta^2\to0$ limit.

The precise width of this plateau depends on parton form factors as well as on parton distributions, 
and should be calculated numerically. However, a rough semi-quantitative estimate of the value of the 
critical transverse momentum $p_0$  where the spectrum starts to flatten out can be obtained 
from the following simplified consideration.

In the double logarithmic approximation one can neglect effects due to parton distributions and state that the plateau develops 
when the product of relevant Sudakov form factors just compensates the pole enhancement:
\[
 \frac{d}{d\ln \kappa^2} \ln \bigg(  \frac{1}{\kappa^2} \prod_i S_i(Q^2,\kappa^2) \bigg) \>=\> 0.
\]
From Eqs.~\eqref{eq:Sudakovs} we obtain the condition
\beq
\Sigma_C  \, \frac{\as(p_0^2)}{\pi} \ln \frac{Q}{p_0} \>=\> 1,
\eeq
where $\Sigma_C$ is the sum of the ``squared color charges'' (Casimir operators) of participating partons. 
Substituting the one-loop expression for the running coupling, we have
\beq\label{eq:4}
 \ln\frac{p_0^2}{\LQCD^2} =  \ln\frac{Q^2}{\LQCD^2} \cdot \left(1+\frac{\beta_2}{2\Sigma_C}\right)^{-1},
\eeq
with $\beta_2=11N_c/3-2n_f/3$ the leading coefficient of the QCD $\beta$-function.
The estimate of the ``flattening momentum'' follows: 
\beq\label{eq:5}
p_0^2 \propto \LQCD^2 \left(\frac{Q^2}{\LQCD^2}\right)^{\gamma}, \quad \gamma =  \left(1+\frac{\beta_2}{2\Sigma_C}\right)^{-1}.
\eeq
Using $\beta_2=9$ for $n_f=3$ light quark flavors, the exponent $\gamma$ in Eq.~\eqref{eq:5} equals $0.372$ for a quark pair 
($\Sigma_C=8/3$), $0.456$ for a system of a quark and a gluon ($\Sigma_C=13/3$), and $0.571$ for two gluons ($\Sigma_C=6$). 

The value $p_0$ characterizes a typical total transverse momentum of the final state produced in a hard 2-parton collision, 
be it a Drell--Yan pair, $W/Z$ or a system of four hadron jets (or, say, three jets and a large transverse momentum photon).

In the case of four- (three-) parton collisions, one has {\em double}\/ back-to-back enhancement, 
accompanied by the {\em four}\/ Sudakov form factors, in place of two. 
So, the pattern of the spectrum flattening in the $\delta^2, \delta'^2 \to 0$ limit remains roughly the same. 

\medskip
We conclude the discussion of the differential distribution with two remarks. 

As we have seen above, in the back-to-back kinematics 
the 4-jet production in double parton collisions ($4\to4$, $3\to4$), 
and due to higher order QCD effects in standard two-parton collisions ($2\to4$), are comparable.  
This makes the experimental studies of double hard collisions not an easy task.
To extract MPI, one should learn to reliably subtract the $2\to4$ contributions. 
This ``background'' should be theoretically predicted and generated at the NLO level (for the amplitude) 
in order to accommodate loop correction effects discussed above in Section~\ref{Sec24}.

On the experimental side, the Tevatron experiments \cite{Tevatron1,Tevatron2,Tevatron3} have set a fashion 
of searching for MPI by looking at {\em angular correlations}\/ between transverse momentum imbalances 
in $\gamma+3\,\mbox{jets}$ (and $\gamma+2\,\mbox{jets}$, \cite{Tevatron3}) events. 
Such a strategy can well be used to signal the presence of MPI. 
At the same time, it is not suited for quantitative analysis as it mixes together contributions of hard and soft physics. 
In order to extract and study double hard collisions, and thus to get hold of inter-parton correlations inside hadrons,   
one should instead base the measurements on the {\em values}\/ (and relative geometry) of transverse momentum imbalances.

\section{Total cross section\label{total}}

Integrating the differential distribution \eqref{eq:sumsigmas} over imbalances $d^2\delta_{13}$, $d^2\delta_{24}$ we obtain the total
cross section of the two dijet production in the double parton collision process in the leading collinear approximation:
\begin{subequations}\label{stot}
\beq
  \frac{d\sigma(x_1,x_2,  x_3,x_4)}{d\hat{t}_1\,d\hat{t}_2}
 = \>\frac{d\sigma^{13}}{d\hat{t}_1}\, \frac{d\sigma^{24}}{d\hat{t}_2} \times \biggl\{\> \frac1{S_4} +  \frac{1}{S_3}\> \biggr\}.
\eeq
Here $S_4$ and $S_3$ are $x_i$- and $Q^2$-dependent parameters that describe effective interaction areas
characterizing 4- and 3-parton collisions, correspondingly.
They are given by the following expressions:
\begin{widetext}
\beq
\begin{split}
S_4^{-1} & (x_1,x_2,x_3,x_4;Q^2)  = \int \frac{d^2\Delta}{(2\pi)^2} \>\>\biggl\{  {}_{[2]}\!D_{a}(x_1,x_2;Q^2,Q^2;\vec\Delta)\,
 { } _{[2]}\!D_{b}(x_3,x_4;Q^2,Q^2;-\vec\Delta) \cr
 &+\,{ }
_{[2]}\!D_{a}(x_1,x_2;Q^2,Q^2;\vec\Delta)\,  _{[1]}\!D_{b}(x_3,x_4;Q^2,Q^2;-\vec\Delta)  \>+
{} _{[1]}\!D_{a}(x_1,x_2;Q^2,Q^2;\vec\Delta)\,  _{[2]}\!D_{b}(x_3,x_4;Q^2,Q^2;-\vec\Delta)  \biggr\},
\label{s4}
\end{split}
\eeq
            and
\beq
\begin{split}
S_3^{-1}(x_1,x_2,x_3,x_4;Q^2) & = \>   \sum_c  \int\frac{d^2 {\Delta}}{(2\pi)^2} P_c^{1,2}\left(\frac{x_1}{x_1+x_2}\right)
 \int^{Q^2} \frac{d\delta^2}{\delta^2}\frac{\alpha_s(\delta^2)}{2\pi} \prod_{i=1}^4 S_i(Q^2,\delta^2) \cr
&\times
\frac{G^c_a(x_1+x_2,\delta^2,Q^2_0)}{x_1+x_2}\>  { } _{[2]}\!D^{3,4}_b(x_3,x_4;\delta^2,\delta^2;\vec\Delta)
\quad +\>\> (a\leftrightarrow b; 1,2 \leftrightarrow 3,4)
\label{s3}
\end{split}
\eeq
\end{widetext}
\end{subequations}
Unlike the case of $4\to4$, at the level of the differential spectrum the contribution $3\to4$ due to an interplay
of non-perturbative and perturbative two-parton correlations is not given by an expression with full derivatives over two relevant scales;
therefore the final expression for $S_3^{-1}$ contains logarithmic integration over the hard scale $\delta^2$
running up to the overall hardness scale of the parton scattering $Q^2$.

\section{Conclusions and Outlook}
\label{Conclude}

In this work we aimed at generalization of the QCD factorization theorem to multiple parton interactions in
hadron--hadron collisions.
Such generalization is necessary for carrying out systematic studies --- both theoretical and experimental ---
of the parton correlations inside hadrons, and allows one to incorporate in a model independent way higher order pQCD effects.
In the small-$x$ domain, generalization of the MPI analysis presented here may be looked upon as
realization of the Gribov Pomeron Calculus, with exact account of energy-momentum conservation. 

We considered four jet production and have shown that double hard parton collisions play a significant role in the {\em back-to-back kinematics}.
This kinematics one selects by demanding the transverse momentum imbalances of two pairs of jets,
$\delta_{13}^2= (\vec{j}_{\perp 1}+\vec{j}_{\perp 3})^2$ and  $\delta_{24}^2= (\vec{j}_{\perp 2}+\vec{j}_{\perp 4})^2$,
to be much smaller than the overall hard scale of the scattering process: $\delta_{13}^2, \delta_{24}^2 \ll Q^2 \sim j_{\perp i}^2$.

We were able to analyze and calculate corresponding differential spectra and total cross sections with account of all collinear enhanced 
pQCD radiative effects giving rise to Sudakov form factors and to scaling violations in one-parton and {\em generalized double parton distributions},
${}_{2}$GPDs, that have been introduced in \cite{BDFS1}.
Our main results are Eqs.~\eqref{eq:DDT4}--\eqref{eq:DDT32} for the differential distribution and Eq.~\eqref{stot} for the total cross section 
of 4-jet production in the back-to-back kinematics.
They are derived in the leading logarithmic (collinear) approximation.
The formalism developed here provides the framework, and a starting point, for systematic calculation of NLL and NLO corrections
that may be needed for more careful account of the MPI effects at LHC.

In particular, in the present study we did not hunt for single logarithmic corrections due to {\em large-angle soft gluon radiation}.
These (formally subleading but potentially important) logarithmic effects depend on details of the color transfers in hard parton scattering processes.
They should be treated together with a delicate question of algorithmic determination of transverse momenta of the hadron jets.
The corresponding study \cite{ABC} exists for the single hard scattering case, $2\to2$, and provides a solid base for analogous treatment 
of large-angle gluon radiation in double parton collision processes.

\medskip

We demonstrated that, in addition to the QCD-improved parton model picture of 4-parton collisions,
pQCD also reserves an important place to 3-parton collisions producing four jets.
Such $3\to4$ transitions are the novel hard double scattering processes that are systematically treated
in this work for the first time.

$3\to4$ parton process appears when a parton from one colliding hadron splits into two partons that enter into double hard interaction 
with two partons from the (non-perturbative) wave function of the second hadron.
Perturbative parton splitting occurs at relatively small impact parameters, while the relative impact parameter distance between 
two hard vertices in a $4\to4$ collision can be as large as the transverse size of the hadron.
Parametrically, $3\to4$ and $4\to4$ subprocesses are comparable (they are of the same order in $\as$ and have similar structure 
of the back-to-back enhancement).
At the same time, the difference in geometry of the interaction processes results in a {\em numerical enhancement}\/ of $3\to4$ 
with respect to $4\to4$ subprocess (by a numerical factor of 4--5).
As a result,  $3\to4$ transitions should contribute significantly to 4-jet production, even if they happen to constitute formally a small correction to a $_2$GPD.

It is important to stress that the processes where one parton splits perturbatively into two in {\em both colliding hadrons}\/ 
do not contribute to the 2-dijet production, since the corresponding differential distribution lacks the necessary back-to-back 
double pole enhancement ${d\sigma}/{(d^2\delta)^2} \propto \delta^{-4}$.
In fact, such configurations represent a one-loop correction to the $2\to4$ jet production mechanism and as such belong 
to the class of the standard single hard interaction processes rather than to MPI.
This result of \cite{BDFS1} has been later confirmed in Refs.~\cite{DiehlSchafer,stirling1}.

\medskip
The role of $3\to4$ parton subprocesses is twofold.

When the perturbative splitting occurs at momentum scales smaller than the pair jet imbalances $\delta^2$
(that is, ``deep inside the DGLAP ladder''), such contributions exhibit the same two-pole enhanced structure as the
conventional $4\to4$ transitions already discussed in \cite{BDFS1}:
\[
 \frac{d\sigma}{d^2\delta_{13}\,d^2\delta_{24}} \>\propto\> \frac{1}{\delta_{13}^2}\cdot \frac{1}{\delta_{24}^2}.
\]
The corresponding differential distribution is a direct generalization of the ``DDT formula'' \cite{DDT} 
containing derivatives over transverse momentum imbalances, with Eq.~\eqref{eq:DDT4} describing $4\to4$ 
collisions and Eq.~\eqref{eq:DDT31} --- a crosstalk between large- and small-distance two-parton correlators.

At the same time, the parton splitting may occur just at the scale $\delta^2$ and produce two partons that,
without any further evolution, interact with two partons from the wave function of the second hadron.
Such eventuality also has a necessary back-to-back enhancement but of a different structure:
\[
 \frac{d\sigma}{d^2\delta_{13}\,d^2\delta_{24}} \propto \frac{1}{\delta'^2}\cdot \frac{1}{\delta^2}, \>
 \delta'^2=\!\! \bigl(\sum_{i=1}^4 \vec{j}_{\perp i}\bigr)^2 \! \ll \delta^2\!=\delta_{13}^2\simeq \delta_{24}^2.
\]
Such contributions are absent in the parton model. They cannot be represented as the product of two $_2$GPDs and
must be taken into account separately, see Eq.~\eqref{eq:DDT32}.
A detailed numerical investigation of these contributions will be presented elsewhere \cite{BDFS2}.

Formulas derived in this work can be used to address various aspects of the LHC physics.
Our results are directly applicable to the case of double Drell--Yan processes at LHC.
In particular they can be applied to the recently observed production of heavy quarkonia at the LHC \cite{LHCb}.
The formalism developed in the present paper should help to improve the accuracy of the prediction of QCD
backgrounds in searches for new physics.
It also can be used for further studies of the nucleon structure,
in particular of the short-range non-perturbative inter-parton correlations in the nucleon and the nuclei,
for the study of the higher twist contributions that are unavailable in the conventional DIS processes, etc.
In addition, it would be interesting to extend the formalism developed here to investigate QCD medium effects that
manifest themselves in $AA$ collisions in recent experiments at CERN.
\smallskip

On the experimental side, observation of the parton splitting processes discussed in this paper
will give the first direct evidence of an interplay between large- and short-distance QCD correlations in hard processes.
To achieve this goal one has to look for enhancement of 4-jet production in the kinematical region of jet transverse momenta
\[
(\vec{j}_{1\perp}+\vec{j}_{2\perp}+\vec{j}_{3\perp}+\vec{j}_{4\perp})^2 \ll (\vec{j}_{1\perp}+\vec{j}_{3\perp})^2\simeq (\vec{j}_{2\perp}+\vec{j}_{4\perp})^2.
\]

\section*{Acknowledgments}
The authors are grateful to the Galileo Galilei Institute for Theoretical Physics, Florence, for an opportunity to get together 
and fine tune this work during the {\em QCD after the start of LHC}\/ Workshop in September--October 2011.

\section*{Appendix}

\appendix
\setcounter{equation}{0}
\renewcommand{\theequation}{A.\arabic{equation}}%

Here we present the analysis of real parton production in the course of $1\to2$ parton splitting, $0\to 1+2$, 
determining the correlated 4-jet production. We consider three contributing diagrams: that of Fig.~\ref{Fig34-2}(a) for
 emission off the initial parton "0", and those of Fig.~\ref{Fig34-1} for emission off the offspring partons "1" and "2".

\subsection{Vertices}
It is straightforward to write down exact expressions for the amplitude using the economic technique of quasi-parton states
 developed by Bukhvostov, Frolov, Lipatov and Kuraev (BFLK; not to be confused with BFKL) and described
 in \cite{BFLK}. In the BFLK technique the quark and (axial gauge) gluon Green
 functions are replaced by propagators of two physical $\pm$ helicity states.
In general, in this framework additional terms looking like contact four-parton
 interaction vertices arise. They, however, do not produce collinear enhanced contributions.

Effective three-parton interaction vertices are proportional to the linear combination of the transverse momenta 
of participating partons weighted with the longitudinal light-cone momentum fractions:
\beq
  K_\perp^\mu = \alpha_i k_{j\perp}^\mu - \alpha_j k_{i\perp}^\mu.
\eeq
In the matrix element the vector index is contracted with the gluon polarization vector (for details see \cite{BFLK}).

The products of two vertices for our amplitudes read
\begin{subequations}
\beeq
0\!\!  &:&\!\!  T_0 \cdot (x_1+x_2+\alpha){\delta'} \times \left( {x_2}{\delta_{13}}- x_1\delta_{24} \right)  \\
1\!\!  &:&\!\!  T_1 \cdot \left(- (x_1\!+\!\alpha){\delta_{13}}-  {x_1} {\delta_{24}} \right)  \!\times\!  (x_1\!+\!x_2){\delta_{24}}  \\
 2\!\!  &:&\!\!  T_2 \cdot  \left( ({x_2+\alpha}){\delta_{24}} + {x_2}{\delta_{13}} \right)  \times  (x_1\!+\!x_2){\delta_{13}}
\eeeq
\end{subequations}
Here the first vector corresponds to the real parton (soft gluon) radiation, the second vertex --- to the internal hard splitting.

\subsection{Amplitudes}
Adding the virtual propagators
\[
 \frac{\alpha}{\delta'^2}, \quad \frac{x_2}{\delta_{24}^2} \quad\mbox{and}\quad  \frac{x_1}{\delta_{13}^2},
\]
correspondingly, and extracting the factor $(x_1+x_2+\alpha)(x_1+x_2)$ we get the amplitudes
\begin{subequations}\label{eq:3amps}
\beeq
\label{eq:3ampsa}
0\!\! &=& T_0 \> \cdot \frac{\alpha \delta' }{\delta'^2}\times \frac{ {x_2}{\delta_{13}}-x_1\delta_{24}}{x_1+x_2}  \cdot P\\
\label{eq:3ampsb}
1\!\!  &=& T_1 \cdot \frac{ -x_1\delta'-\alpha{\delta_{13}}}{x_1+x_2+\alpha}  \times\frac{x_2\delta_{24}}{\delta_{24}^2}  \cdot P \\
\label{eq:3ampsc}
2\!\!  &=& T_2 \cdot \frac{ x_2\delta'+\alpha {\delta_{24}}} {x_1+x_2+\alpha}  \times\frac{x_1\delta_{13}}{\delta_{13}^2} \cdot P.
\eeeq
\end{subequations}
Here $P$ is the common propagator factor defined in \eqref{eq:Pdef}:
\beq\label{eq:rdef}
   P^{-1} = x_1\delta_{24}^2 + x_2\delta_{13}^2 + (x_1+x_2)\,r^{-1} \,\delta'^2, \quad   r \> \equiv \>  \frac{\alpha(x_1+x_2)}{x_1x_2}.
\eeq

\subsection{Limits}
In the case of soft gluon radiation, $\alpha\ll x_i$, the parameter $r$ defined in \eqref{eq:rdef} becomes numerically small.
Therefore one should examine, separately, two kinematical regions:
\smallskip

\noindent
{\bf 1.} $\displaystyle \delta'^2 \ll r\cdot \delta^2$

\noindent
Only "0" contributes:
\begin{subequations}
\beq\label{eq:0}
 M \>\simeq\> T_0 \>\> \cdot \frac{\alpha \delta' }{\delta'^2}\times \frac{ \delta_{13}}{(x_1+x_2)\delta_{13}^2}
\eeq

\bigskip
\noindent
{\bf 2.} $\displaystyle \delta^2 \gg \delta'^2 \gg r\cdot \delta^2$

\noindent
Here the amplitudes \eqref{eq:3ampsb} and \eqref{eq:3ampsc}  contribute, and we obtain
\beq\label{12}
M\>\simeq\>(T_1+T_2)\cdot \frac{x_1x_2\,\delta'}{x_1+x_2}\, \frac{\delta_{13}}{\delta_{13}^2} \>\frac{\alpha}{x_1x_2\,\delta'^2}
\eeq
\end{subequations}
reproducing the structure of \eqref{eq:0}.
Emissions from 1 and 2 are coherent, and due to conservation of the color current, $T_1+T_2=T_0$, the two expressions coincide.
As a result, \eqref{eq:0} applies in the entire kinematical region $\displaystyle \delta'^2 \ll \delta^2$, irrespectively to the value of $r$.

This contribution --- real parton production off the incoming line "0" --- is a part of the $3\to 4$ process, Eq.~\eqref{eq:DDT32}, 
described in Section~\ref{Sec:324}.

\bigskip
Additional collinear enhanced contributions arise in two complementary regions.
\smallskip

\noindent
{\bf 3.} $\displaystyle \delta_{24}^2\ll \delta_{13}^2\simeq \delta'^2$

\noindent
Here the amplitude 1 dominates:
\begin{subequations}\label{eq:ap12}
\beq
  M \>\simeq\> T_1\cdot \frac{\alpha}{x_1+x_2+\alpha}\>\frac{\delta_{13}}{\delta_{13}^2}\frac{\delta_{24}}{\delta_{24}^2}.
\eeq

\bigskip
\noindent
{\bf 4.} $\displaystyle \delta_{13}^2\ll \delta_{24}^2\simeq \delta'^2$

\noindent
Alternatively, here the dominant amplitude is that with emission off the line 2:
\beq
  M \>\simeq\> T_2\cdot \frac{\alpha}{x_1+x_2+\alpha}\>\frac{\delta_{13}}{\delta_{13}^2}\frac{\delta_{24}}{\delta_{24}^2}.
\eeq
\end{subequations}
The contributions \eqref{eq:ap12} are related with internal evolution of the $_2$GPD of the hadron $a$
and are contained in Eq.~\eqref{eq:DDT4}.
Specifically, in the kinematical region where the  imbalances of two jet pairs are significantly different, these contributions correspond to
the case when the differentiation over the smallest of the two imbalances applies to the $_2$GPD and acts upon the integration limit
of the correlation ${}_{[1]}D$ term in  \eqref{eq:termPT}.

%
%

\end{document}